\documentclass[aps,showpacs,twocolumn]{revtex4-1}
\usepackage{epsfig}
\usepackage{nicefrac}
\usepackage{bm}
\usepackage{dcolumn}
\usepackage{graphics}
\usepackage{amsmath}
\usepackage{amssymb}
\usepackage{color}
\newcommand{\bra}{\langle}
\newcommand{\ket}{\rangle}

\newcommand{\bs}[1]{\ensuremath{\boldsymbol{#1}}}
\newcommand{\red}[1]{{#1}}

\newcommand{\be}{\begin{equation}}
\newcommand{\ee}{\end{equation}}
\newcommand{\beqa}{\begin{eqnarray}}
\newcommand{\eeqa}{\end{eqnarray}}
\newcommand{\bk}{\bs k}
\newcommand{\ev}[1] {|\bra #1  \ket |^2}
\begin{document}
\title{Multipole Analysis of Radio-Frequency Reactions in Ultra-Cold Atoms}
\author{Betzalel Bazak, Evgeny Liverts and Nir Barnea}
\date{\today}
\affiliation{The Racah Institute of Physics, The Hebrew University, 91904, Jerusalem, Israel}

\begin{abstract}
  Using the multipole expansion we analyze photo induced reactions in an ultra-cold 
  atomic gas composed of identical neutral bosons.
  While the Frank-Condon factor dominates the photo induced spin-flip
  reactions, we have found that for frozen-spin process where the atomic spins
  are conserved the reaction rate  
  is governed by the monopole $r^2$ and the quadrupole terms.
  Consequently, the dependence of the frozen-spin reaction rate on the photon wave number
  $k$ acquires an extra $k^4$ factor in comparison to the spin-flip process.  
  Comparing the relative strength of the $r^2$ and quadrupole modes in
  dimer photoassociation we predict that 
  the mutual importance of these two modes changes with temperature and scattering length.
\end{abstract}

\pacs{25.10.+s,25.20.-x,67.85.-d,31.15.ac}
\maketitle


Radio-Frequency (rf) association of molecules in ultra-cold atomic traps
is an invaluable experimental tool for studying the universal properties of 
few-body systems \cite{GreenePT2010, Braaten06}.
In such experiments 
\cite{Thompson05, Rapp06, Weber08, Lompe10, Lev10, Lev11, Nakajima11, Lev12} 
rf induced dimer or trimer formation leads to enhanced atom loss rate due to larger 
probability for three-body recombination.
The measured atom loss rate as a function of the rf field frequency
incorporates the data about the various molecular thresholds and structures,
and can be used to calibrate the two-body scattering length and to measure the 
three-body binding energy. 

Photo-integration of molecules, where resonant rf radiation is used to stimulate 
photon emission, is closely related, through time reversal, to electro-weak 
reactions in light nuclei. In fact, rf photoreactions inducing an atomic 
spin-flip are  governed by the Frank-Condon factor. This factor is the 
atomic analog of the Fermi operator, which at low momentum 
transfer describes the charged vector-current neutrino induced isospin-flip 
reactions in 
nuclear systems. The nuclear analog to frozen-spin
rf reactions is photoabsorption, where an incoming photon dissociates the nucleus. 
 {For example, photoreactions within the $\alpha$-cluster model of nuclei 
such as $^{12}$C, $^{16}$O are governed by the same physical mechanisms studied here.}
Therefore, studying rf photoreactions in ultra-cold atomic systems can be
instrumental for understanding the nuclear dynamics, the complicated nuclear
continuum, 
and for predicting unmeasurable electro-weak cross-sections of  
astrophysical importance.

Previous analysis of these rf experiments \cite{Chin05, Bertelsen06, Hanna07, Klempt08},
have relied on the Franck-Condon factor estimating the transition rates. While
this approach is appropriate for describing spin-flip reactions, it is
not suitable for frozen-spin reactions.
In \cite{Trimer} we have studied the quadrupole contribution to the frozen-spin 
response of a bound bosonic trimer.
In this article we present a complete multipole analysis of 
the rf association process forming a bound molecule composed of $N$ identical bosons.
We show that the spin-flip and frozen-spin processes differ by their operator
structure and by the de-excitation modes that contribute to the
photoassociation rate. We apply our results to study frozen-spin dimer formation. 
We focus on bosonic alkali-metal atoms, in which a single $s$-shell
valence electron is coupled to an half-integer spin nucleus to form a boson.  
This is the case for $^7$Li, $^{23}$Na, $^{39}$K, $^{41}$K, $^{85}$Rb,
$^{87}$Rb, and $^{133}$Cs. Extension of our results to other atoms and
fermionic isotopes is rather simple but lays outside the confines of the
current manuscript. 

In ultra-cold atomic traps, the trapped particles are subject to an external
magnetic field $\bs B=B\hat z$ applied to control the 
inter-atomic interaction, i.e. the scattering length, through a Feshbach resonance. 
Consequently, the energy of the system is dominated by the interactions of the
valence electron spin $S=\nicefrac{1}{2}$ with the external and the nuclear magnetic fields.
The Zeeman interaction with the external magnetic field reads
$-g \mu_B \bs S \cdot \bs B$, 
where 
$g$ is the electron g-factor, and $\mu_B$ is the Bohr magneton \cite{Units}.
The hyperfine spin-nucleus contact interaction 
$\hbar \omega_{hf} \bs I \cdot \bs S$
couples the electron with the nuclear spin  $\bs I$
where $\omega_{hf}$ is the hyperfine splitting frequency.
When both interactions are comparable, the total spin $\bs F=\bs I+\bs S$ is
not a good quantum number,
however its 
projection $m_F$, is.


{To be specific,} we consider $I=\nicefrac{3}{2}$, 
the case for $^7$Li, $^{23}$Na, $^{39}$K, $^{41}$K, and $^{87}$Rb.
The atomic spin state composition in the $|m_S,m_I \ket$ basis,
$$|m_F\ket=\sin\theta_{m_F}|\nicefrac{1}{2},m_F-\nicefrac{1}{2}\ket
+\cos\theta_{m_F}|\nicefrac{-1}{2},m_F+\nicefrac{1}{2}\ket
$$
is given by 
\be \label{Zeeman}
\tan \theta_{m_F}=\\
\begin{cases}
\infty & m_F=2\\
(1-2\eta-2\sqrt{1-\eta+\eta^2})/\sqrt{3}   & m_F=1\\
-\eta-\sqrt{1+\eta^2}  & m_F=0\\
-(1+2\eta+2\sqrt{1+\eta+\eta^2})/\sqrt{3}  & m_F=-1 \\
\end{cases} \ee
for the low energy branch states.
Here 
$\eta=g\mu_BB/2\hbar\omega_{hf}$ 
is the ratio between the Zeeman and the hyper-fine splitting.


In photoassociation experiments, an rf radiation of few MHz is applied to the
ultra-cold atomic gas.
When the photon energy matches, stimulated emission occurs 
and a molecule is formed. 
Fermi's golden rule dictates the transition rate of
such process to be,
\be \label{golden_rule}
r_{i\rightarrow f}=\frac{2\pi}{\hbar} \bar{\sum_i} \sum_f
|\bra f, \bk \rho | \hat{H}_I | i\ket|^2 
\delta(E_i-E_f-\hbar\omega_k) \;,
\ee 
where $\bar\sum_i$ averages on the appropriate initial states and $\sum_f$ sums
on the final states. Specifically, $|i\ket$ is the continuum state, $\bk
\rho$ is the emitted photon momentum and polarization, and $|f\ket$ is the
formed $N$-body bound state.  

The interaction Hamiltonian 
$\hat{H}_I=-\frac{e}{c}\int d\bs x \bs J({\bs x})\cdot \bs A(\bs x)$
between an atomic system and an electromagnetic (EM) radiation
field $\bs A$ is defined by the current operator $\bs J$
composed of convection and magnetization terms, 
$\bs J (\bs x)= \bs J_c(\bs x)+\red c \nabla \times \bs \mu(\bs x)$.
For a system of neutral atoms interacting with long wavelength radiation
$\bs J_c\approx 0$,
and the interaction between the radiation field and the system 
occurs through the magnetization density 
$e\bs\mu(\bs x)=g \mu_B \sum_i\bs S_i \delta(\bs x-\bs r_i)$.
Using box normalization of volume $\Omega$, the EM field reads 
$
\bs A(\bs x)=\sum_{k,\rho} \sqrt{\frac{\hbar c^2}{2\Omega\omega_k}}\hat e_{\bk \rho}
(a^\dagger_{\bk \rho}e^{i\bs k \cdot \bs x}+h.c.)
$
where $\rho=1,2$ are the photon linear polarizations,
$\omega_k$ is its frequency and $\bk$ is its momentum.

For an atomic system we can assume that the initial and final matter
wave functions can be written as 
a product of spin and configuration space terms, $\Psi=\Phi_{L M_L}\chi_{M_F}$,
where $\Phi_{L M_L}=\Phi_{L M_L}(\bs{r}_1,\bs{r}_2,\ldots \bs{r}_A)$ is a symmetric
$N$-particle wave function with angular momentum quantum numbers $L M_L$
and $\chi_{M_F}={\cal S}\left[|m_F(1)\ket|m_F(2)\ket\ldots | m_F(A)\ket\right]$ is the 
symmetrized spin wave function with magnetic quantum number $M_F=\sum_j m_F(j)$. 
Using this factorization, 
The transition matrix element in (\ref{golden_rule}) takes the form
\begin{align}
\bra f ,\bk \rho| \hat H_I | i \ket=
-ig\mu_B\sqrt{\frac{\hbar \red {c^2}}{2\Omega\omega_k}} &
\sum_{j=1}^N \cr
\bra \chi^f_{M'_F} | \bs S_j \cdot (\bk \times \hat e_{\bk \rho}) |
\chi^i_{M_F} \ket 
      \bra & \Phi^f_{ L' M_L'} | e^{i\bs k \cdot \bs r_j}| \Phi^i_{ L
  M_L} \ket \;.
\end{align}

The spin operator can be written in spherical notation,
$
\bs S \cdot (\bk \times \hat e_{\bk \rho})=\sum_\lambda
(-)^{\lambda}S_{-\lambda} \cdot (\bk \times \hat e_{\bk \rho})_\lambda 
$
where $\lambda=0,\pm 1$.
To calculate this quantity the
laboratory $\hat z$ axis is defined by the static magnetic field. The
rotation from the laboratory reference frame $(\hat x,\hat y,\hat z)$, to the
photon reference frame, $(\hat e, \hat k \times \hat e, \hat k)$ is
represented by Euler angles $(\alpha,\beta,\gamma)$.
Then, 
$(\hat k \times \hat e)\cdot \hat z = - \sin \gamma \sin \beta$, and
$|(\hat k \times \hat e)_{\pm 1}|^2 = (1-\sin ^2 \gamma \sin^2  \beta)/2$.

For rf radiation the photon wavelength is much larger then the typical dimension of the
system $R$, therefore $kR\ll1$ and  the lowest order in $kR$ dominates the interaction. 
The first contribution depends on the photon energy versus the energy difference between
adjacent Zeeman states. 

Using Eq. (\ref{Zeeman}), the spin matrix elements can be easily calculated to give,
\be
\bra m_F'|S_\lambda|m_F\ket=\delta_{m_F+\lambda,m_F'}
\begin{cases} 
\sin \theta_{m_F'} \cos \theta_{m_F}   & \lambda=+1 \\
-\frac{1}{2}\cos 2\theta_{m_F}         & \lambda= 0 \\
\sin \theta_{m_F} \cos \theta_{m_F'}   & \lambda=-1
\end{cases} 
\ee
If the photon can induce a Zeeman state change, i.e. spin-flip, the leading contribution
comes at order $k$. Energy is delivered to the system
through the spin matrix element and we can approximate $e^{i x}\simeq 1$, 
to get 
\begin{align}\label{spin-flip}
&\ev{f ,\bk \rho| \hat H_I | i}=
\frac{g^2\mu_B^2\hbar \red c k(1-\sin ^2 \gamma \sin^2 \beta)N^2}{4\Omega}
\\ \nonumber
&\sum_{\lambda=\pm 1}\ev{\chi^f_{M'_F} |S_{j,\lambda}| \chi^i_{M_F}}
\ev {\Phi^f_{ L' M_L'} | \Phi^i_{ L M_L} }
 \delta_{L,L'} \delta_{M_L,M_L'} \;.
\end{align}
The factor $N^2$ results from utilizing the symmetry of the wave function and
$S_{j,\lambda}$ is the magnetic moment operator of {any specific} particle $j$. The 
last term on the rhs of Eq. (\ref{spin-flip}) is just the Frank-Condon
factor. 

In frozen spin reactions $\lambda=0$ and the rf photon does not affect the spin structure of
the system. Therefore $M_F'=M_F$, $|\chi^f_{M_F}\ket = | \chi^i_{M_F}\ket $ and
the transition matrix element can be written as
\begin{align} \label{frozen-spin}
\bra f,\bk \rho | \hat H_I | i\ket =& -i g \mu_B \sqrt{\frac{\hbar \red {c^2}}{2\Omega\omega_k}} k \sin \gamma \sin \beta \bra S_{0}  \ket \cr &
\bra \Phi^f_{L' M_L'} | \sum_{j=1}^N e^{i\bs k \cdot \bs r_j} | \Phi^i_{L M_L} \ket \;,
\end{align}
where 
$\bra S_{0} \ket = \bra \chi^i_{M_F} | S_{1,0} | \chi^i_{M_F}\ket $ 
is the average single particle magnetic moment.

In the long wavelength limit, the leading contributions to the photoreaction
process are 
\begin{align} \label{expand}
   \sum_{j=1}^N e^{i\bs k \cdot \bs r_j} &\approx  N
 +i\sum_{j=1}^N \bk \cdot \bs r_j
 - \sum_{j=1}^N  \frac{k^2 r_j^2}{6} \cr &
 - \sum_{j=1}^N {4\pi\frac{k^2 r_j^2}{15} \sum_m Y_{2 -m}(\hat{k}) Y_{2 m}(\hat{r}_j)},  
\end{align}
where $Y_{lm}$ are the spherical harmonics.
The difference between the four operators in (\ref{expand}) is clear. The first operator is
proportional to 1 and stands for elastic interaction. For
photon emission this kind of interaction defies energy conservation.  
The second term is just the dipole operator, which for identical particles is
proportional to the center of mass and hence does not \red{affect relative motion of the atoms}. Therefore the only contributions to 
\red {such} excitation arises from the $r^2$, and the quadrupole
terms. Consequently the transition probability scales as $k^5$. 
Summing over the initial and final magnetic numbers $M_L$, $M'_L$, 
the transition matrix element reads 
\begin{align} \label{t-frozen}
&{{\displaystyle \sum_{M_L}}}\sum_{M_L'}\ev{f,\bk \rho | \hat H_I | i}  = 
\frac{4\pi \hbar \red c k^5 g^2 \mu_B^2 \sin^2\gamma \sin^2 \beta N^2}{2\Omega}
 \ev{S_0} \cr &
\left(
\frac{1}{6^2} \ev{\Phi^f_{L'} \Vert r_j^2 Y_0\Vert \Phi^i_{L}}
+\frac{1}{15^2} \ev{\Phi^f_{L'} \Vert r_j^2 Y_2(\hat r_j)  \Vert \Phi^i_{L}}\right).
\end{align}
Here, like in (\ref{spin-flip}), we have used the symmetry of the wave function
and $\bs{r}_j$ is the coordinate of {any of the particles $j=1,\ldots N$}.
Comparing equations (\ref{t-frozen}) and  (\ref{spin-flip}) the difference between
spin-flip and frozen-spin reactions is evident. To begin with, the
frozen-spin acquires an extra $k^4$ factor suppressing the reaction rate.
Then, we see that whereas the leading configuration space spin-flip operator
is proportional to the unit operator, like the Fermi operator in nuclear
physics, for frozen-spin reactions we identify two competing modes of order
$k^2$, one is the $r^2$ operator and the other is just the quadrupole operator.


Once established the operator structure of the transition rate we would like
to focus our attention on frozen-spin photoassociation reactions trying to
estimate the relative importance of the two reaction modes.
To this end we consider the case of dimer formation, 
which under certain simplifying assumptions can be calculated analytically.

In the universal regime, where the scattering length is much larger than
the potential range, one can approximate the wave functions by their
asymptotic behavior.  
The dimer bound state with binding energy $E_B$ is an $s$-wave state, and
$\Phi^f_{00}(\bs r)=Y_{00}\,\varphi_B(r)/r$ 
where 
$\bs{r}=\bs{r}_2-\bs{r}_1$ is the relative coordinate.
$\varphi_B(r)\cong\sqrt{2\kappa}e^{-\kappa r}$ 
and
$\kappa=\sqrt {m |E_B|/\hbar^2}$. 
For the initial (continuum) states, the asymptotic scattering solution is used, i.e. 
$\Phi^i_{LM}(\bs r)=Y_{LM}\,\phi_L(r)/r$, 
where
$\phi_L(r)=\sqrt{2/R}qr[j_L(qr)\cos \delta_L - y_L(qr)\sin \delta_L ]$, 
$\delta_L$ is the phase shift,
$j_L$ and $y_L$ are the spherical Bessel functions of the first and second
kind, $q$ is the relative momentum, and the wave function is normalized in a 
sphere of radius $R$.

Given the $L'=0$ dimer ground state, the $r^2$ term corresponds to 
$L=0$ $s$-wave formation, while the quadrupole term to $L=2$ $d$-wave formation. 
Using $\ev{0 \Vert Y_{0} \Vert 0}=1/4\pi$
and 
$\ev{2 \Vert Y_{2} \Vert 0}=5/4\pi$, 
\beqa \label{rme-sd}
\ev{\Phi^f_0 \Vert r^2 Y_0\Vert \Phi^i_0}&=&\frac{1}{4\pi} I_s^2
\cr
\ev{\Phi^f_0 \Vert r^2 Y_2 \Vert \Phi^i_2} &=&\frac{5}{4\pi} I_d^2.	
\eeqa
Here 
$I_s=\int_0^\infty \varphi_B^{*}(r) \phi_0(r) r^2 dr,$
and same for $I_d$ replacing $\phi_0$ by $\phi_2$.
The transformation to the relative coordinate $r$ in (\ref{rme-sd}) adds an
extra factor of $1/16$ in (\ref{t-frozen}).

For the $s$-wave mode, the radial integration yields
\be
I_s=
\frac{4q}{(q^2+\kappa^2)^3} \sqrt{\frac{\kappa}{R}}
\left[(3\kappa^2-q^2)\cos \delta_0
-\frac{\kappa}{q}(3q^2-\kappa^2)\sin \delta_0 \right].
\ee
At low scattering energy, the $s$-wave phase shift takes the form
$ q \cot \delta_0\cong-\nicefrac{1}{a_s}+r_{eff}q^2/2$
where $a_s$ is the scattering length and $r_{eff}$ is the effective range. 
The dimer's binding energy is connected to these parameters by the relation \cite{LL3}
$\kappa=\nicefrac{1}{a_s}+{r_{eff}}\kappa^2/2 $.

For the $d$-wave mode, the radial integration yields,
\beqa
I_d &= &\frac{2q}{(q^2+\kappa^2)^3}\sqrt{\frac{\kappa}{R}}
\left[ 8q^2 \cos \delta_2 \right. \cr & & \left.
+
\frac{\kappa}{q^3}(15q^4+10q^2\kappa^2+3\kappa^4)\sin \delta_2 \right] \;.
\eeqa
For short-range potentials, $\delta_L\approx q^{2L+1}$ as $q \longrightarrow 0$, thus
$ \delta_2 \approx (a_d q)^5 $
where $a_d$ is the $d$-wave ``scattering length''. 
We note that for the Van der Waals potential the correct threshold behavior is, 
$ \delta_2\approx\pm(a_d q)^4 $ \cite{Levy63}. As long as $a_d$ is of the 
order of the effective range it is safe to assume $\delta_2\approx 0$.

\begin{figure}\begin{center}
\includegraphics[width=8.6cm,trim=0cm 0cm 0cm 0mm,clip=true]{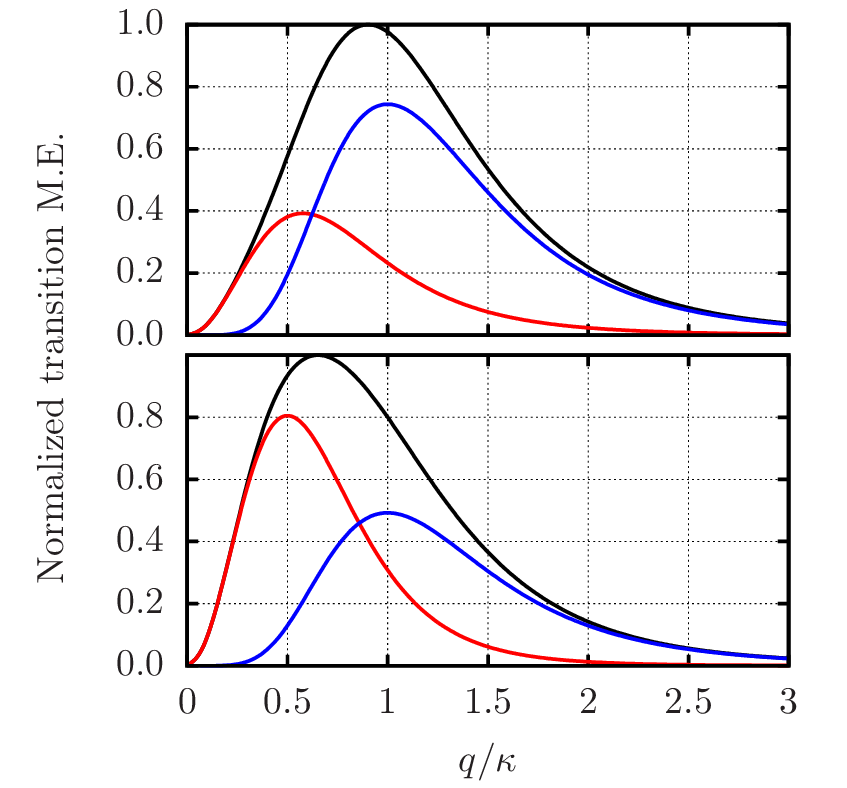}
\caption{\label{Fig:Uni} (Color online) The normalized 2-body transition matrix
  element, Eq. (\ref{t-frozen}), as a function of the relative momentum
  $q/\kappa$.
  The sum (black), the $r^2$ (red, peaked at
  $q=\kappa/2$) and quadrupole (blue, peaked at $q=\kappa$) terms are given for
  $a/r_{eff}=2$ (upper panel) and  $a/r_{eff}=200$ (lower panel). }
\end{center}
\end{figure}

The relative contribution of the $s,d$ modes to the dimer formation is displayed
in  Fig. \ref{Fig:Uni}, where the last term in parenthesis on the rhs of
Eq. (\ref{t-frozen}) is presented normalized, along with the s and d components. 
From the figure it can be seen that the $s$-wave association is peaked around 
$q=\kappa /2$, while the $d$-wave association is peaked at $q=\kappa$. 
For small $a_s/r_{eff}$ ratio
the $d$-wave is the dominant mode but as
we approach the unitary limit $a_s/r_{eff}\longrightarrow \infty$ the $s$-wave
becomes dominant. {We note however, that as the $a_s/r_{eff}$ ratio become
  smaller finite range corrections, ignored here, become more important.}
In photoassociation  experiments
angular analysis is impossible and therefore the different energy dependence
of the two dimer formation modes, Fig. \ref{Fig:Uni}, can be used to distinguish 
between them.
As it stands, the large overlap between the two modes
will certainly complicate such an attempt.

The average of the initial states $\bar\sum_i$ and the sum over the final states $\sum_f$
are the last components needed 
for evaluating the dimer association rate, Eq. (\ref{golden_rule}).
The initial two-body state $|i\ket=|qLM_L\ket$ describes two atoms in the
continuum with relative momentum $q$ and angular momentum $L, M_L$.
The average on these states takes the form
\be \label{sumi}
\bar{\sum_i}=\sum_{L=0}^\infty \sum_{M_L=-L}^{L} \int_0^\infty dq P(q L M_L) \;,
\ee
where $P(q L M_L)$ is the probability of finding an atomic pair in an internal
state $|q L M_L\ket$ due to thermal distribution.
We assume that the system is in thermal equilibrium with temperature $T > T_{BEC}$
higher then the condensation temperature. 
The partition function of a pair confined in a sphere of
radius $R$ is given by \cite{LLV} 
\begin{align}
\mathcal{Z} =&\sum_{n \in bound \;\;states} e^{-\beta E_n}
\cr
+& \frac{1}{\pi} \int_0^\infty dq \sum_{L=0}^{L_{max}} (2L+1) (R+\frac{d\delta_L}{dq}) e^{-\beta \hbar^2 q^2 / m} \;.
\end{align}
Given $q$, finite $R$ imposes a cutoff $L_{max}\approx qR$ on $L$.
For $L=0,2$ and large enough $R$ we can safely assume that
$ P(qLM_L) = P(q) = \frac{1}{\mathcal{Z}}\frac{R}{\pi}e^{-\beta \hbar^2 q^2 / m}. $

The sum $\sum_f$ contains all possible dimer
quantum numbers and
all possible photons weighted by the emission function, 
\be \label{sumf}
\sum_f=\frac{\Omega}{(2\pi)^3}\sum_{\rho,L',M'_L}\int d\bk ({1+N_{\bk\rho}}),
\ee
where $N_{\bk\rho}$ is the number of photons with
momentum $\bk$ and polarization $\rho$ in the initial state.
The stimulating rf radiation is a narrow distribution centered at some 
$\bk_{rf}$. Therefore 
$N_{\bk\rho} \approx N_{\bk_{rf}\rho_{rf}}\delta(\bk-\bk_{rf})\delta_{\rho,\rho_{rf}}/\Omega$,
where $\bk_{rf}$ is the momentum of the stimulating rf field
and $\rho_{rf}$ is its polarization.

Substituting Eqs. (\ref{t-frozen}), (\ref{sumi}), (\ref{sumf}) into Fermi's golden
rule (\ref{golden_rule}), the dimer formation rate is given by
\begin{align}
r_{i\rightarrow f}=&\frac{g^2 \mu_B^2 \red c k_{rf}^5 m}{16 (2\pi)^2\hbar^2 q} 
\frac{N_{\bk_{rf},\rho_{rf}}}{\Omega}P(q)
\cr
&\ev{S_0}\sin^2\gamma\sin^2\beta
\left ( \frac{1}{6^2}I_s^2+\frac{5}{15^2}I_d^2 \right ).
\end{align}
The relative momentum $q$ is connected to the photon wave number through energy conservation 
$ \hbar^2 q^2/m=E_B+\hbar c k_{rf}$.

\begin{figure}\begin{center}
\includegraphics[width=8.6cm,trim=0cm 0cm 0cm 0mm,clip=true]{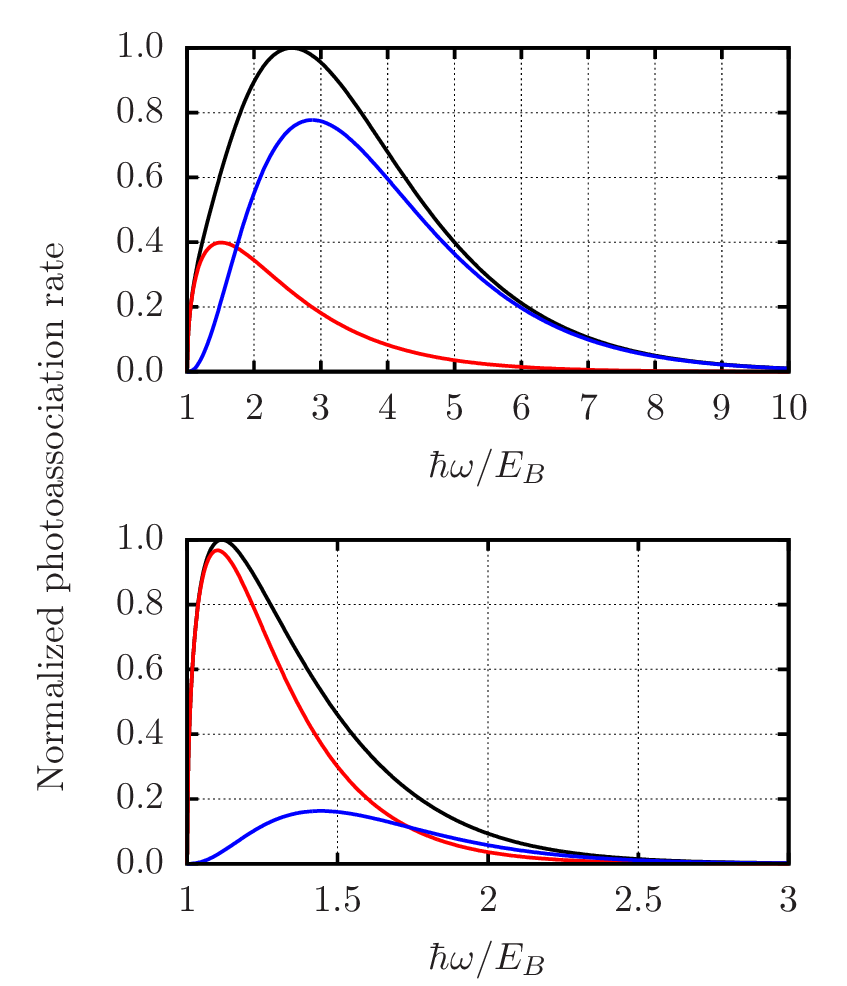}
\caption{\label{fig:rates} (Color online) Normalized dimer photoassociation rate
as a function of rf photon frequency. The total (black), the $r^2$ (red,
low energy peak) and the quadrupole (blue) rates are presented for $^7$Li with
 $a_s=1000a_0$ at $T=5\mu\rm{K}$ (lower panel) and at $T=25\mu\rm{K}$ (upper panel).} 
\end{center}
\end{figure}

\begin{figure}\begin{center}
\includegraphics[width=8.6cm,trim=0cm 0cm 0cm 0mm,clip=true]{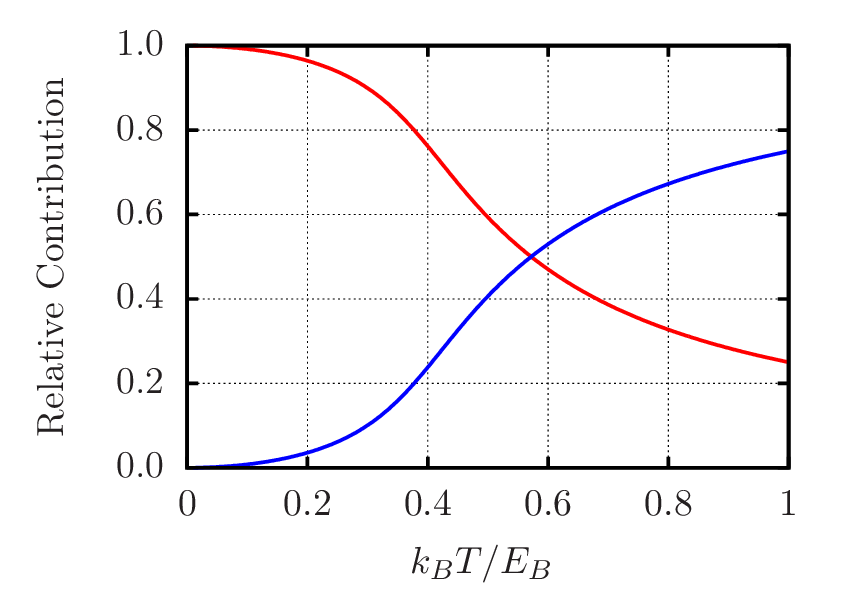}
\caption{\label{fig:peakratio} (Color online) The relative contribution of the
  $s$ (red, light grey) and $d$ modes to the resonance peak
  as a function of temperature
  $k_BT/E_B$. }
\end{center}
\end{figure}

The relative importance of $s,d$ modes shifts with temperature. This point is
demonstrated in Figs. \ref{fig:rates} and \ref{fig:peakratio}. 
In Fig. \ref{fig:rates} we present the $s,d$ rates
to the photoassociation of $^7$Li dimers at $a=1000 a_0$, $a_0$ is the Bohr radius,
for $T=5\mu\rm{K}$ and $T=25\mu\rm{K}$.
The relative  contribution of the $r^2$ and
quadrupole to the resonance peak is displayed in Fig. \ref{fig:peakratio}.
It is evident that at low temperatures the $s$ mode is dominant. The imprtance
of the $d$ mode grows with the ratio $k_B T/E_B$. Therefore we can conclude that for small $k_B
T/E_B$ values the photoassociation is an $s$-wave process while for large
values it is a $d$-wave process.

Summing up, in this work we have studied the photo response of ultra-cold
atomic gases. 
Comparing spin-flip and frozen-spin experiments we have found that the
reaction mechanism is quite different in these two processes. 
While the spin-flip reaction is dominated by the well known Frank-Condon
factor, for frozen-spin process the reaction rate  
  is governed by the two competing $r^2$ and quadrupole modes.
  Comparing the relative strength of these modes in
  dimer photoassociation we predict that 
  the mutual importance of these two modes changes with temperature and scattering length.
The implications of these results on photoassociation experiments and on
trimer formation rates is yet to be studied.

\section*{ACKNOWLEDGEMENTS}
This work was supported by the ISRAEL SCIENCE FOUNDATION 
(Grant No.~954/09).
The authors would like to thank L. Khaykovich, S. Jochim, and N. Nevo Dinur 
for their useful comments and suggestions during the preparation of this work.


\begin{thebibliography}{00}
\bibitem{Units} We adopt the Heaviside-Lorentz units in this paper.
\end{thebibliography}
\end{document}